\documentclass{jps-cp}
\usepackage{txfonts}
\usepackage{color}
\usepackage{bm}
\usepackage{graphicx}
\usepackage{soul}
\usepackage{upgreek}

\title{CeCu$_{2}$Si$_{2}$ and YbRh$_{2}$Si$_{2}$: Strange Cases of Heavy-Fermion Superconductivity}

\author{Z.~Y.~Shan$^{1}$, M.~Smidman$^{1,2}$, O.~Stockert$^{3}$, Y.~Liu$^{1,2}$, H.~Q.~Yuan$^{1,2,4,5}$, P.~J.~Sun$^{6}$, S.~Wirth$^{3}$, E.~Schuberth$^{7}$, and F.~Steglich$^{3,1}$\thanks{steglich@cpfs.mpg.de}}
\inst{$^1$Center for Correlated Matter and School of Physics, Zhejiang University, Hangzhou 310058, China \\
$^2$Zhejiang Province Key Laboratory of Quantum Technology and Device, Department of Physics, Zhejiang University, Hangzhou 310058, China \\
$^3$Max Planck Institute for Chemical Physics of Solids, 01187 Dresden, Germany \\
$^4$State Key Laboratory of Silicon Materials, Zhejiang University, Hangzhou 310058, China \\
$^5$Collaborative Innovation Center of Advanced Microstructures, Nanjing University, Nanjing 210093, China \\
$^6$Institute of Physics, Chinese Academy of Sciences, Beijing 100190, China \\
$^7$Physics Department, Technical University of Munich, 80333 Munich, Germany} 

\recdate{July 28, 2022}

\abst{The heavy-fermion superconductor CeCu$_{2}$Si$_{2}$ exhibits two-band, $d$-wave superconductivity with a finite energy gap over the whole Fermi surface around the magnetic instability where 4$f$ antiferrerromagnetic order is suppressed. In contrast, in YbRh$_{2}$Si$_{2}$ heavy-fermion superconductivity appears only when 4$f$-electronic antiferromagnetic order is replaced at ultra-low temperatures by a combined nuclear and 4$f$-spin order. Whereas both compounds exhibit different variants of antiferromagnetic instabilities, i.e., a spin-density-wave quantum critical point in CeCu$_{2}$Si$_{2}$ and one of “partial-Mott” type in YbRh$_{2}$Si$_{2}$, in both cases the Cooper pairing, as well as the pronounced ``strange-metal'' behavior in YbRh$_{2}$Si$_{2}$, appear to be driven by large-to-small Fermi surface fluctuations. The transport properties and scanning tunneling spectroscopy (STS) for these materials are dominated by single-ion Kondo scatterings down to very low temperatures. Further open problems of the Kondo lattice include both the interplay between superconductivity and antiferromagnetic order as well as the onset of lattice coherence. While microscopic coexistence of superconductivity and antiferromagnetism seems to require a sufficiently large staggered moment, the onset of lattice coherence in transport measurements and STS is associated solely with the crystal-field-doublet ground state, while it involves the fully degenerate Hund's rule multiplet in ARPES.}

\begin{document}
\maketitle

\section{Kondo-lattice coherence and quantum criticality}

Though in the focus of condensed-matter physicists for several decades \cite{Cornut_1972}, Kondo-lattice systems are still fascinating targets of contemporary basic research. These are usually certain intermetallic compounds of the rare earths Ce, Sm and Yb where extremely strong Coulomb correlations within the localized $4f$-shells are instrumental to form huge effective masses of the charge carriers (``heavy fermions'', HFs) and give rise to distinct low-temperature properties, such as heavy Fermi liquid, HF superconductivity (SC), spin-density-wave (SDW) order and quantum criticality. In this article, we discuss new insights into some aspects of Kondo-lattice behavior as well as HF SC of the isostructural tetragonal materials CeCu$_{2}$Si$_{2}$ and YbRh$_{2}$Si$_{2}$.

\begin{figure}[t]
	\begin{center}
		\includegraphics[width=0.7\columnwidth]{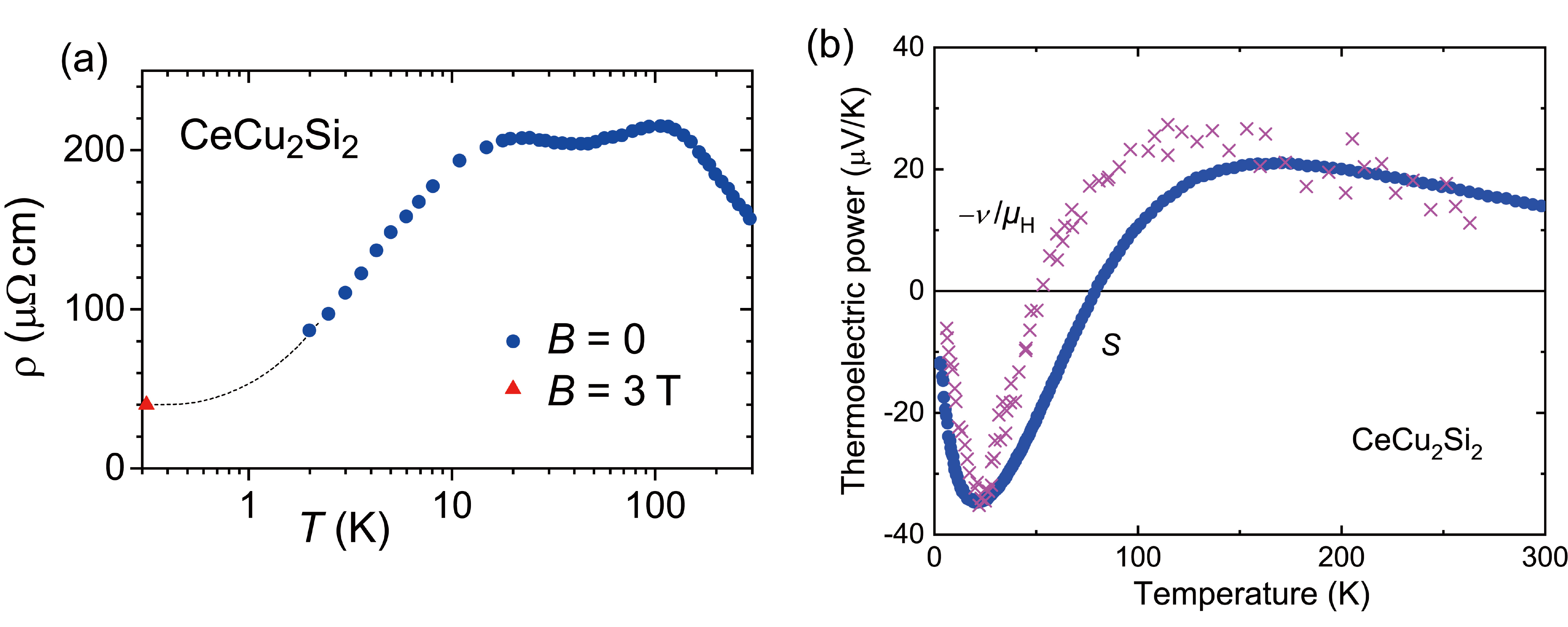}
	\end{center}
	\caption{Transport properties of CeCu$_2$Si$_2$. (a) Electrical resistivity $\rho$ vs $T$ on a logarithmic scale, registered in a home-built $^{4}$He cryostat. Clearly, there is no saturation in $\rho(T)$ at the lowest temperature $T \approx 1.5$~K. In order to determine the residual resistivity $\rho_{0}$, the sample was mounted into a self-built $^{3}$He evaporator, equipped with a superconducting magnet. As explained by a footnote in Ref. \cite{Franz_1978}, $\rho(T)$ turned out to suddenly fall to zero at around 0.5~K. To suppress this apparent superconducting state and determine $\rho_{0}$, a magnetic field as high as $\approx$ 3~T had to be applied, see red triangle. Replotted from Franz \textit{et al.}, Z. Phys. B $\bm{31}$, 7 (1978) [Ref. \cite{Franz_1978}]. (b) Temperature dependence of the thermoelectric power $S$ vs $T$ and $-\nu$/$\mu_{\rm H}$ vs $T$ for CeCu$_2$Si$_2$ measured up to 300~K, $\nu$: Nernst coefficient, $\mu_{\rm H}$: Hall mobility. Adapted with permission from Sun and Steglich, Phys. Rev. Lett. $\bm{110}$, 216408 (2013) [Ref. \cite{Sun_2013}]. Copyright 2013 by the American Physical Society.}
	
	\label{fig1}
\end{figure}

As displayed in  Fig.~\ref{fig1}a, the temperature dependence of the electrical resistivity of CeCu$_{2}$Si$_{2}$ displays a double-peak structure, with a logarithmic temperature dependence at high temperatures \cite{Franz_1978,Ocko_2001}. The latter indicates single-ion Kondo scattering off the fully degenerate $J = 5/2$ Hund's rule multiplet of Ce$^{3+}$, its position in energy being labeled k$_{\rm B}T_{\rm K}^{\rm high}$ \cite{Cornut_1972}. The decrease of $\rho(T)$ below $T_{\rm coh}\approx15$~K indicates the onset of \textit{coherent} Kondo scattering, clearly implying that the mean-free path of the charge carriers is much larger than the interatomic spacing (Mott-Ioffe-Regel value). $T_{\rm coh}$ is frequently found to agree well with $T_{\rm K}$, the Kondo temperature referring to the crystal-field-derived doublet ground state where the molar entropy assumes Rln2. As shown in  Fig.~\ref{fig1}b, a corresponding temperature dependence is observed for the thermoelectric power (TEP) $S(T)$ \cite{Sun_2013}. By using Mott's formula, for CeCu$_{2}$Si$_{2}$ the temperature dependence of its TEP can be written as $S(T) = -\nu/\mu_{\rm H} - (3e)^{-1}\uppi^{2} {\rm k_{\rm B}}^{2} T \partial {\rm ln}N/\partial \varepsilon \big|_{\varepsilon = \varepsilon_{\rm F}}$ \cite{Sun_2013}. Here, $N(\varepsilon)$ is the conduction-band density of states, whereas $-\nu/\mu_{\rm H}$ is the ``scattering term'', with $\nu(T)$ the Nernst coefficient and $\mu_{\rm H}(T)$ the Hall mobility, determined by both the skew-scattering-derived anomalous Hall coefficient and the magnetic contribution to the resistivity, $\mu_{\rm H} = R_{\rm H}^{\rm a}/\rho_{\rm mag}$. As shown in Fig.~\ref{fig1}b, $S(T)$ is well reproduced, in a very wide temperature window from (at least) room temperature to way below $T_{\rm coh}$, by $-\nu(T)/\mu_{\rm H}(T)$. This observation of dominating conduction-electron scattering off the single Ce$^{3+}$ (Kondo) ions implies that $S(T)$ may be explained by the $renormalized ~HF ~band ~structure ~of ~the ~Kondo ~lattice$ only at very low $T$ \cite{Zlatic_2007}. It supports a successful theoretical treatment of anomalous Hall-effect results for CeCu$_{2}$Si$_{2}$\cite{Aliev_1983} in the framework of a $local$ Fermi-liquid model \cite{Coleman_1985}. Very similarly, based on specific-heat and TEP data obtained for Ce$_{1-x}$La$_{x}$Ni$_{2}$Ge$_{2}$, Pikul \textit{et al.} could demonstrate the importance of $local$ Kondo screening even for the $coherent ~Fermi$-$liquid$ phase in samples with enhanced Ce concentration ($x ~\textless ~0.5$) \cite{Pikul_2012}.

\begin{figure}[t]
	\begin{center}
		\includegraphics[width=0.55\columnwidth]{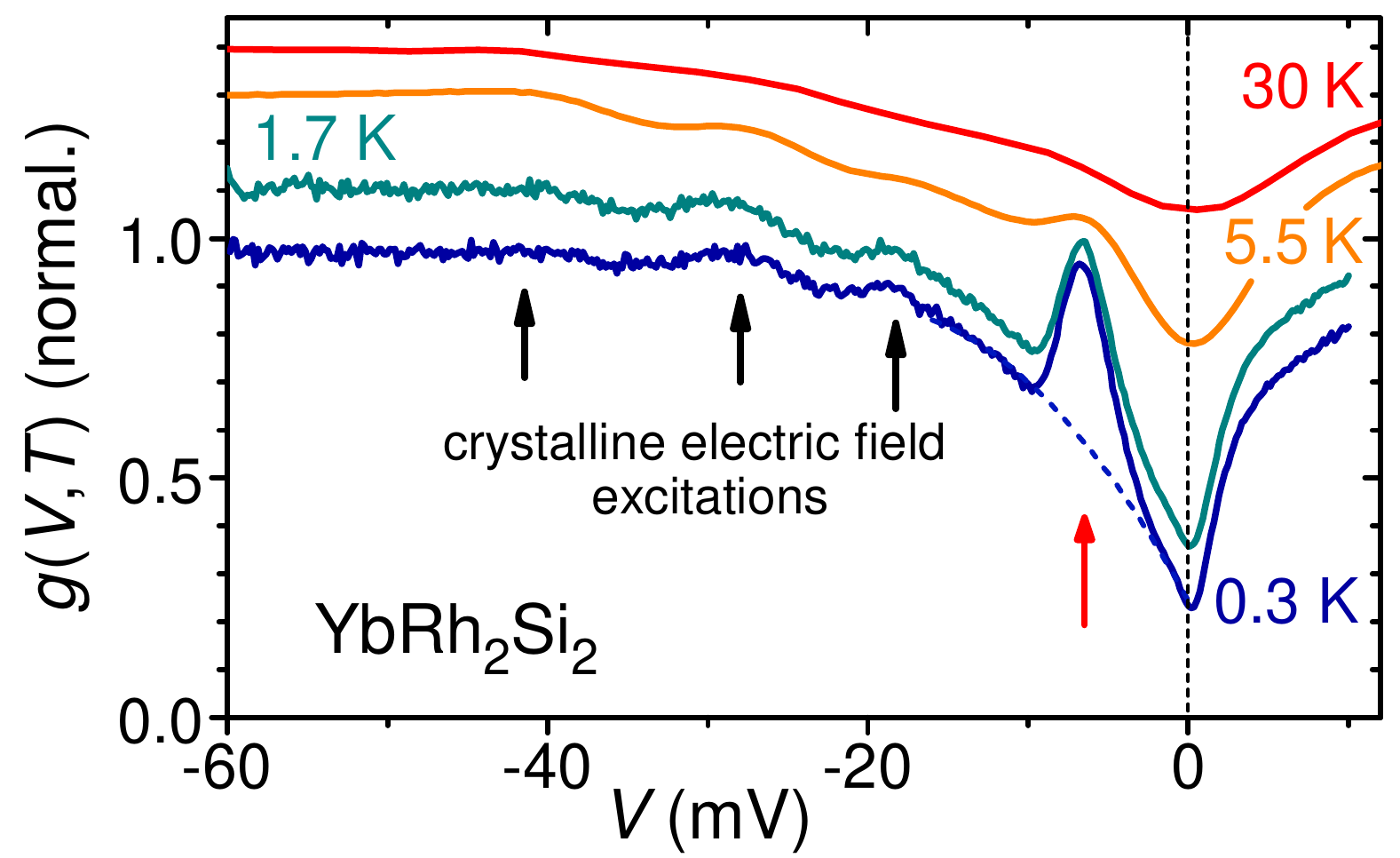}
	\end{center}
	\caption{Scanning tunneling spectroscopy results for YbRh$_2$Si$_2$ obtained at several temperatures up to $T_{\rm coh} \approx 30$~K. Adapted from Seiro \textit{et al.}, Nat. Commun. $\bm{9}$, 3324 (2018) [Ref. \cite{Seiro_2018}], available under a Creative Commons Attribution 4.0 International License. }
	
	\label{fig2}
\end{figure}

For YbRh$_{2}$Si$_{2}$, no low-$T$ peak is resolved in $\rho(T)$ because of a substantial broadening of the CF states; however, this peak can be made visible by weakening the strength of the Kondo effect, i.e., by partially substituting Lu for Yb \cite{Kohler_2008}. In fact, scanning tunneling spectroscopy (STS) into the Si-terminated surface of an YbRh$_{2}$Si$_{2}$ single crystal (ensuring that tunneling takes place into conduction-band states only) revealed a dip at the Fermi energy $\varepsilon_{\rm F}$ that becomes visible, upon cooling, at $T\approx100$~K (close to $T_{\rm K}^{\rm high}$). This indicates a reduction of the conduction-band density of states (DOS) due to the formation of a ``Kondo resonance'' in the local $4f$-DOS at $\varepsilon_{\rm F}$ \cite{Ernst_2011,Seiro_2018}. As seen in Fig.~\ref{fig2}, at $T_{\rm K} \approx 30$~K a weak hump develops at a binding energy of $\approx 6$~meV, indicating the formation of a ``partial hybridization gap'' in the $4f$-states as a consequence of the onset of lattice coherence in the Kondo scattering \cite{Zwicknagl_1992}. Interestingly, while this ``coherence peak'' in STS develops only quite slowly down to about 3~K, i.e., $\approx T_{\rm K}/10$ where it is almost saturating, it starts rising enormously at somewhat lower temperatures. This agrees well with the evolution of the temperature dependence of the TEP coefficient $S(T)/T$ \cite{Hartmann_2010} and suggests that a moderately heavy Fermi-liquid phase, reflecting substantial Kondo-lattice correlations, has to form before distinct low-$T$ phenomena, e.g., quantum criticality and HF SC, can develop. Even at the lowest temperature of the STS experiment (0.3~K) the depth of the dip at $\varepsilon_{\rm F}$ is found to further increase \cite{Seiro_2018}. In line with the TEP data for CeCu$_2$Si$_2$ [Fig.~\ref{fig1}b], this clearly indicates that single-ion scattering processes take place way below the onset of spatial coherence (at $T_{\rm coh} \approx 30$~K), while a \textit{fully} renormalized heavy quasiparticle band structure can be established at substantially lower temperatures only.

\begin{figure}[t]
	\begin{center}
		\includegraphics[width=0.65\columnwidth]{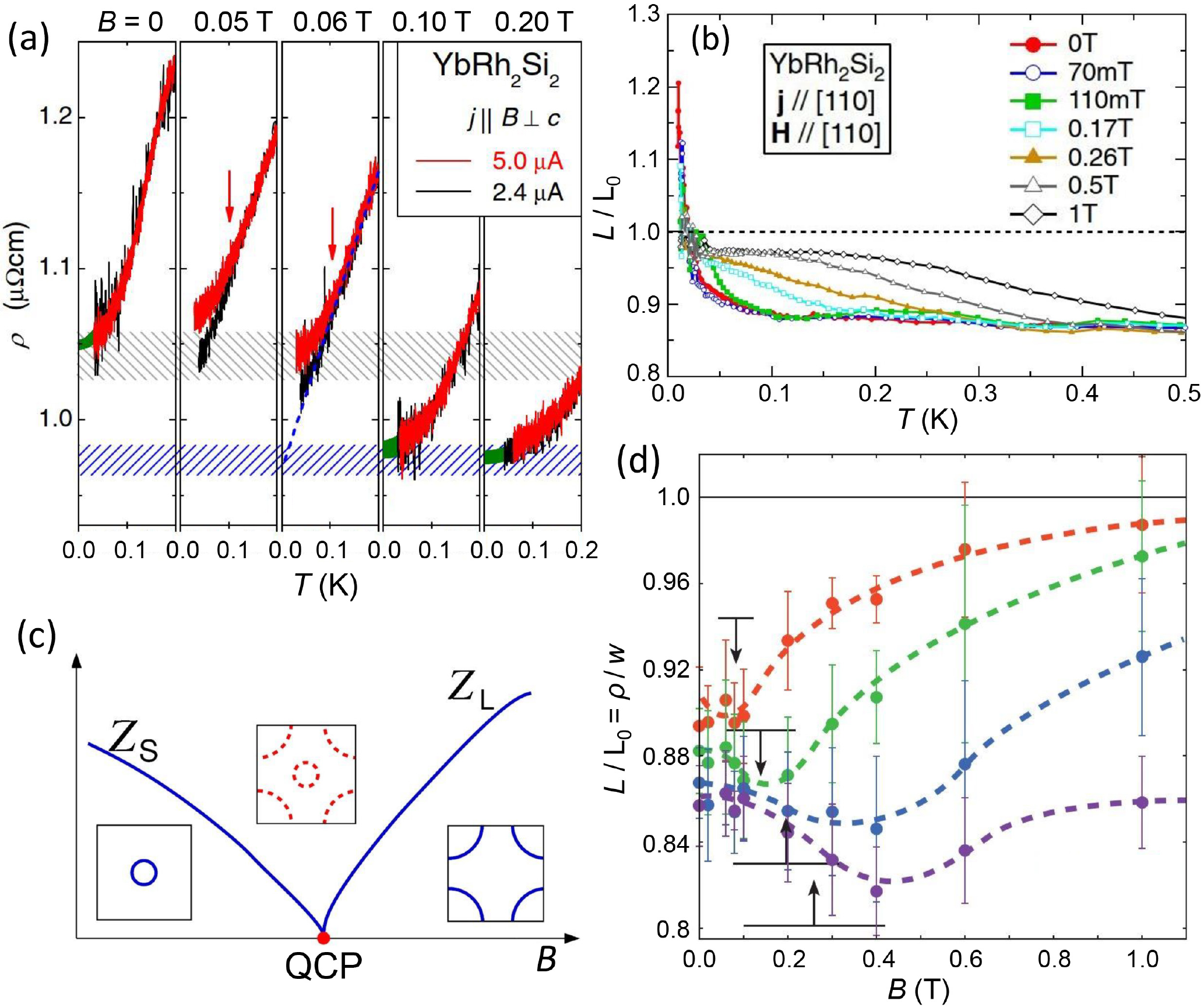}
	\end{center}
	\caption{(a) Low temperature resistivity of YbRh$_{2}$Si$_{2}$ in different applied fields, for both $B ~\textless ~B_{\rm N}$ and $B ~\textgreater ~B_{\rm N}$. Reproduced from Steglich \textit{et al.}, J. Phys. Soc. Jpn. $\bm{83}$, 061001 (2014) [Ref. \cite{Steglich_2014}]. Copyright 2014 by the Physical Society of Japan. (b) Temperature dependence of the Lorenz ratio $L/{\rm L_{0}}$ in different applied fields. Reproduced from Pourret \textit{et al.}, J. Phys. Soc. Jpn. $\bm{83}$, 061002 (2014) [Ref. \cite{Pourret_2014}]. Copyright 2014 by the Physical Society of Japan. (c) Evolution of the quasiparticle weights of both the small ($Z_{\rm S}$) and large ($Z_{\rm L}$) FSs upon crossing a field-induced local QCP. (d) Evolution of the field dependence of $L/{\rm L_{0}}$ at different low temperatures where thermal transport is purely electronic. The red, green, blue and purple data correspond to respective temperatures of 0.1, 0.2, 0.3, and 0.4 K. (c) and (d) are both reproduced from Pfau \textit{et al.}, Nature $\bm{484}$, 493 (2012) [Ref. \cite{Pfau_2012}].}
	
	\label{fig3}
\end{figure}

YbRh$_{2}$Si$_{2}$ is a model system for a (magnetic field-induced) Kondo-destroying AF QCP at $B_{\rm N} \approx 60$~mT ($B \perp c$) where the charge-carrier concentration is expected to abruptly change \cite{Si_2001,Coleman_2001,Si_2010}. This can be directly observed in the results of electrical-resistivity measurements (Fig.~\ref{fig3}a) \cite{Steglich_2014} which show that the residual resistivity registered in the antiferromagnetically ordered phase (with small carrier concentration or ``small Fermi surface (FS)'' in the zero-temperature limit), i.e., at $B ~\textless ~B_{\rm N}$, is indeed substantially larger than at $B ~\textgreater ~B_{\rm N}$ in the paramagnetic phase (with large carrier concentration, or ``large FS''). Also inferred from these data is that close to the QCP, the sample is heated-up by a moderate current, which is ascribed to deteriorated heat transport as a consequence of the violation of the fundamental Wiedemann-Franz (WF) law. The latter strictly holds only for elastic scatterings, notably at $T = 0$. In YbRh$_{2}$Si$_{2}$, at $B = B_{\rm N}$ both the electrical and electronic thermal resistivities, $\rho(T)$ and $w(T) = {\rm L_0}T/\kappa(T)$ [${\rm L_0} = (\uppi {\rm k_B})^2/(3{\rm e}^2)$ being Sommerfeld's constant and $\kappa$ the electronic thermal conductivity], are strictly linear in $T$ over an extended temperature range \cite{Pfau_2012,Pourret_2014}, on approaching their residual values $\rho_0$ and $w_0$ at $T = 0$. The so-called Lorenz ratio $L/{\rm L_0} = \rho\kappa/{\rm L_0}T = \rho_0/w_0$ is found to be $\approx 0.9$ at least up to $T = 0.5$~K as well as in the zero-temperature limit, see Fig. \ref{fig3}b \cite{Pfau_2012,footnote2}; i.e., it is reduced by $\approx$ 10\% compared to the WF-value $L(T = 0)/{\rm L_0} = 1$ \cite{Pfau_2012}. A similar observation was reported for the HF metal YbAgGe \cite{Dong_2013}. As argued in Refs. \cite{Pfau_2012,Schuberth_2022}, this violation of the WF law in YbRh$_{2}$Si$_{2}$ is due to a new kind of electron-electron scattering process, namely between the large and small FSs which coexist at $B_{\rm N}$ even in the zero-temperature limit owing to the smooth disappearance of their corresponding quasiparticle weights at the QCP, see Fig.~\ref{fig3}c \cite{Pfau_2012}. By this scattering not only  momentum but also energy is relaxed, which causes $w_{0}$ to exceed $\rho_{0}$ by $\approx$ 10\%. At finite $T$, the same scattering process apparently results in an unusual minimum of the isothermal control-parameter  dependence of the Lorenz ratio $L(B)/{\rm L_0}$, see Fig.~\ref{fig3}d. It shifts towards $B = B_{\rm N}$, considerably narrows upon cooling (presumably becoming a delta function at $T = 0$ \cite{Pfau_2012}) and highlights an additional inelastic scattering process, along with the occurrence of ``strange-metal'' behavior seen in the resistivities. Because of the very large mean-free path, this $T$-linear dependence of $\rho(T)$ cannot be related to ``Planckian'' dissipation \cite{Bruin_2013,Taupin_2022}. These \textit{fermionic} quantum critical fluctuations between small and large FS are to be distinguished from the aforementioned single-ion Kondo scatterings, which were shown to be of elastic nature \cite{Steglich_1975} and, for Kondo ions with spin $1/2$, have been described in the framework of a local Fermi liquid model \cite{Nozieres_1974}.

Recent ARPES measurements on CeCoIn$_{5}$ have highlighted an interesting new meaning of the energy scale k$_{\rm B}T_{\rm K}^{\rm high}$: They revealed ``hybridized $4f$-bands'' up to $T = 120$~K which cannot be resolved at higher temperatures \cite{Chen_2017}, $cf$. also data on YbRh$_{2}$Si$_{2}$ \cite{Kummer_2015}. Further work is badly needed to explore the apparent dichotomy in the onset of Kondo-lattice coherence, taking place at $T_{\rm K}^{\rm high}$ in ARPES results, compared to $T_{\rm coh} \approx T_{\rm K}$, where Kondo-lattice coherence commences in transport properties and STS.

\begin{figure}[t]
	\begin{center}
		\includegraphics[width=0.7\columnwidth]{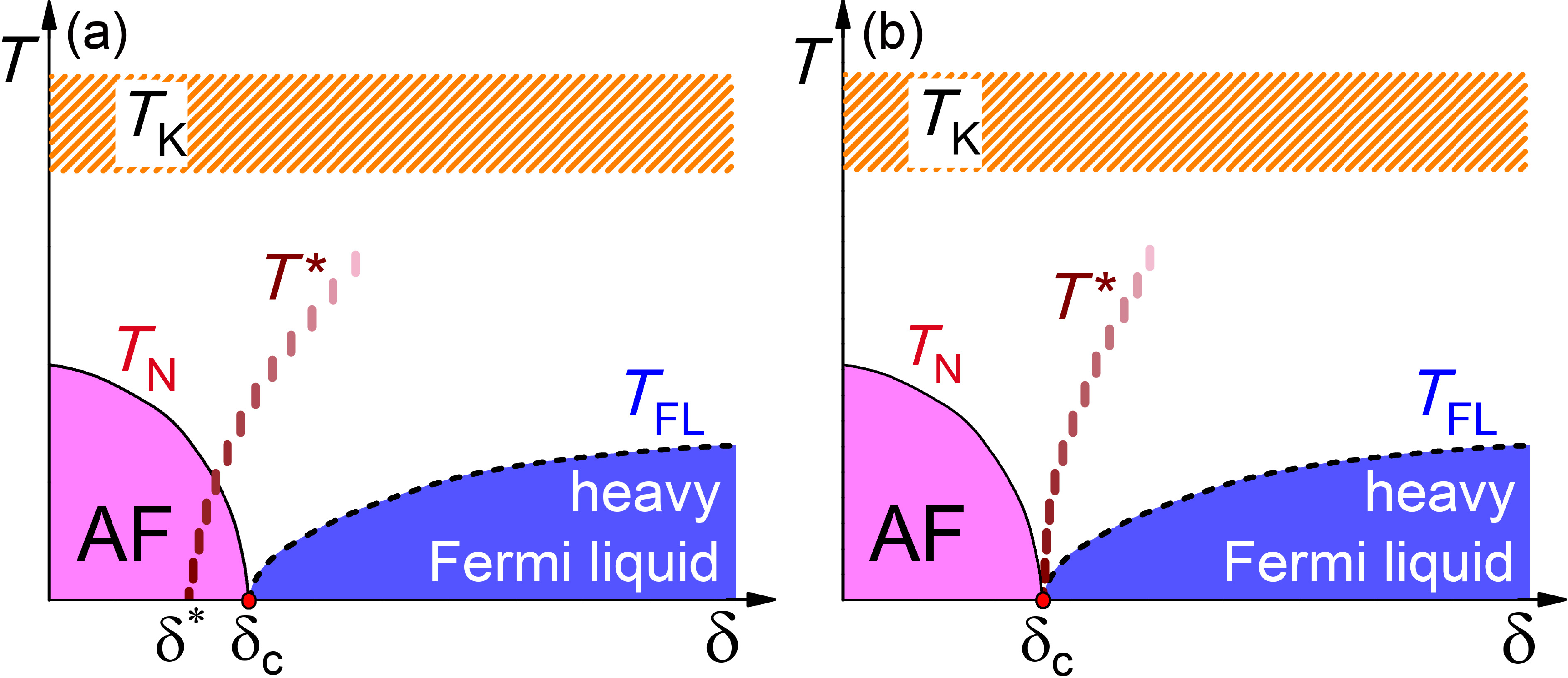}
	\end{center}
	\caption{Illustration of  two possible quantum critical scenarios, where the left corresponds to an itinerant spin-density wave type QCP, while the right is for a local Mott-type QCP. Reproduced from Steglich and Wirth, Rep. Prog. Phys. $\bm{79}$, 084502 (2016) [Ref. \cite{Steglich_2016}]. Copyright 2016 IOP Publishing Ltd.}
	
	\label{fig4}
\end{figure}

Like for other unconventional superconductors, e.g., cuprates, organic transfer salts and Fe-pnictides, SC in HF metals is frequently found in close proximity to magnetic order, often near an AF quantum critical point (QCP) \cite{Stewart_2001and2006}. Here, magnetic order smoothly disappears as a function of a suitable non-thermal control parameter, notably pressure or magnetic field \cite{Lohneysen_2007,Si_2010}. Two variants of HF AF QCPs have been studied intensively in the past, (i) the \textit{itinerant} three-dimensional (3D) SDW-type \cite{Hertz_1976,Millis_1993,Moriya_1995}, (ii) the \textit{local} Mott-type QCP \cite{Si_2001,Coleman_2001} see Figs.~\ref{fig4}a and b \cite{Steglich_2016,Gegenwart_2008}. At the former instability, the heavy charge carriers behave like $d$-electrons in a transition-metal compound, i.e., they stay intact upon tuning through the QCP, while at a \textit{local} QCP these composite charge carriers disintegrate. CeCu$_{2}$Si$_{2}$ is considered to be a prototype quantum-critical metal of type (i) \cite{Arndt_2011,Stockert_2011}, with a low Mott-crossover scale k$_{\rm B}T^{*}$, $T^{*} = (1 - 2)$~K \cite{Smidman_2022}. Since the QCP is located inside the very narrow homogeneity range in the chemical Ce-Cu-Si phase diagram, tiny variations of the Cu/Si ratio enable the growth of \textit{homogeneous} single crystals, which are either superconducting (``$S$-type'', tiny Cu-excess), AF (``$A$-type'', tiny Cu-deficit) or exhibit both SC and AF order (``$A/S$-type'', nearly stoichiometric). YbRh$_{2}$Si$_{2}$ \cite{Custers_2003}, along with CeCu$_{6-x}$Au$_{x}$ \cite{Schroder_2000}, and CeRhIn$_{5}$ \cite{Shishido_2005,Park_2006} are prototypical quantum-critical systems of type (ii). In a third scenario, the AF and Kondo-destroying instabilities are separated by a quantum-critical phase (rather than a QCP). Examples are Ir- and Ge-substituted YbRh$_{2}$Si$_{2}$ \cite{Friedemann_2009,Custers_2010} and CePdAl \cite{Zhao_2019}.

\section{CeCu$_{2}$Si$_{2}$: $d$-wave superconductivity with finite energy gap}

Owing to the strong pair-breaking effect of dilute localized spins in BCS superconductors \cite{Matthias_1958,Abrikosov_and_Gorkov_1961}, the discovery of bulk SC in the Kondo-lattice system CeCu$_{2}$Si$_{2}$ \cite{Steglich_1979} $-$ with one localized $4f$-spin in each unit cell $-$ came as a big surprise and was not generally accepted by the community \cite{Hull_1981,Schneider_1983}. The Cooper pairs in CeCu$_{2}$Si$_{2}$ are formed by extremely heavy quasiparticles, whose effective mass $m^{*}$ exceeds the electron mass, $m_{\rm e}$, by a factor of $\approx 1000$. Its renormalized Fermi velocity ${v_{\rm F}}^{*}$ is thus even smaller than the velocity of sound. The kinetic energy of the heavy charge carriers is of the order ${\rm k_{\rm B}}T_{\rm K}$, with $T_{\rm c}/T_{\rm K} \approx 0.04$ being \textit{much larger} compared to $T_{\rm c}/T_{\rm F} = 10^{-4} - 10^{-3}$ for a conventional BCS superconductor. This highlights CeCu$_{2}$Si$_{2}$ as a ``high-$T_{\rm c}$'' superconductor in a normalized sense. In addition, the ratio $T_{\rm K}/\theta_{\rm D}$ ($\theta_{\rm D}$: Debye temperature) is found to be $\approx 0.5$, which is \textit{much smaller} than  $T_{\rm F}/\theta_{\rm D} \approx 100$, typical for main-group and transition metals. The latter ratio warrants the electron-phonon coupling to be retarded, which efficiently suppresses the onsite Coulomb repulsion between the partner electrons of the Cooper pairs in a BCS superconductor.

Results of neutron-diffraction experiments on $A$-type CeCu$_{2}$Si$_{2}$ have supported \cite{Stockert_2004} earlier conclusions from bulk measurements \cite{Gegenwart_1998} that the AF order is a 3D SDW with small staggered moment, i.e., $\mu_{\rm s} \geq 0.1 \mu_{\rm B}$. The incommensurate propagation wave vector \bm{$Q_{\rm AF}$} was found to be almost identical to the nesting wave vector \bm{$\tau$} connecting the parallel flat regions of the warped part of a cylindrical renormalized (HF) FS oriented along the $c^{*}$ axis, as obtained by renormalized band calculations (RBC) \cite{Zwicknagl_1993}. Inelastic neutron-scattering (INS) measurements performed at $T = 70$~mK and \bm{$q$} = \bm{$Q_{\rm AF}$} have revealed a spin gap that develops at 0.2~meV in the spin-excitation spectrum of superconducting CeCu$_{2}$Si$_{2}$ \cite{Stockert_2011}. The peak at the edge of the spin gap exhibits a positive linear dispersion – characteristic of an acoustic paramagnon \cite{Stockert_2011,Song_2021,Smidman_2022}. As discussed in Ref. \cite{Smidman_2022}, the peak of the magnetic response which occurs inside the superconducting gap of CeCu$_{2}$Si$_{2}$ develops in the one-particle channel, i.e., out of the quasi-elastic line in the normal-state, similar to what was also observed for UPd$_{2}$Al$_{3}$ \cite{Sato_2001} and CeCoIn$_{5}$ \cite{Song_2016}.

\begin{figure}[t]
	\begin{center}
		\includegraphics[width=0.7\columnwidth]{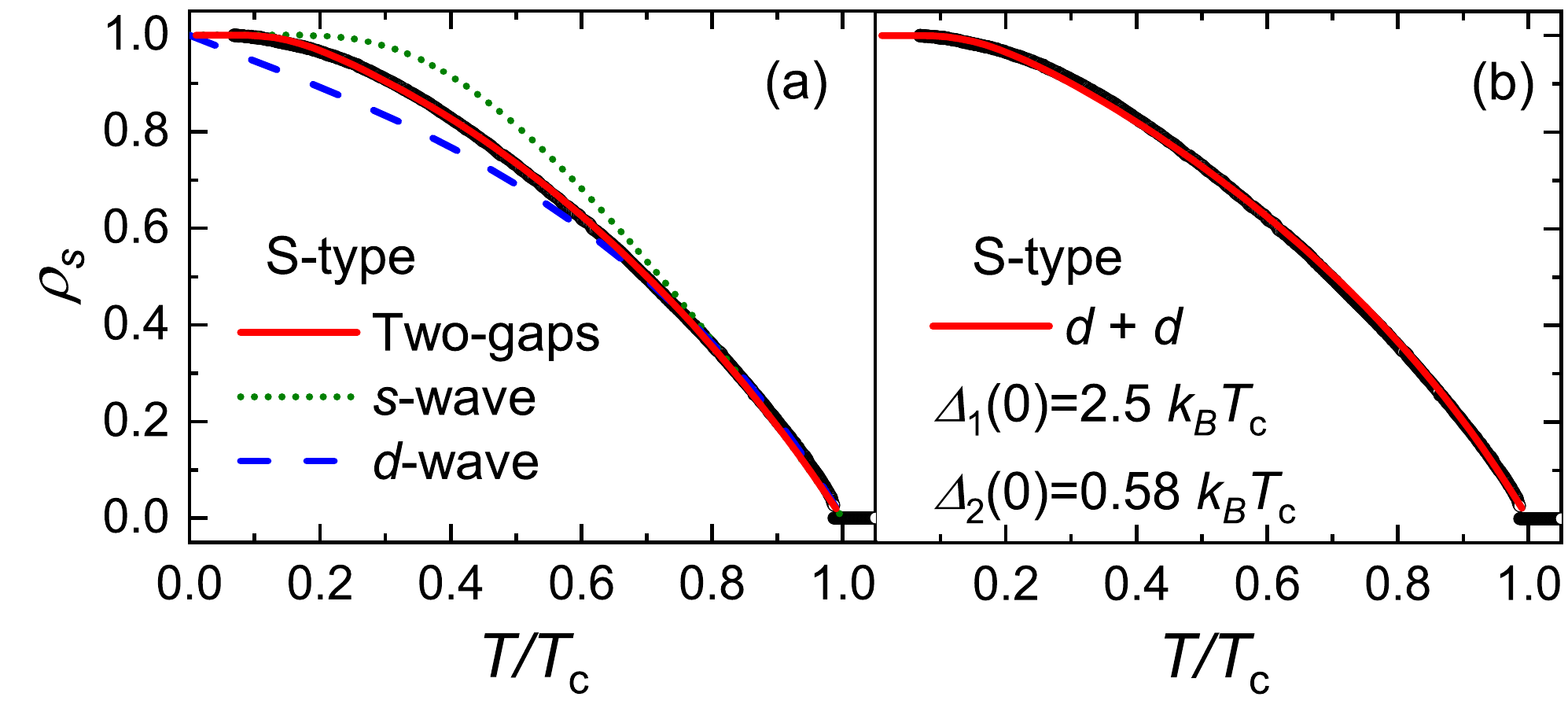}
	\end{center}
	\caption{Temperature dependence of the superfluid density of $S$-type CeCu$_2$Si$_2$ derived from penetration depth measurements using the tunnel-diode oscillator based method. The left and right panels display fits using two isotropic gaps, and a $d + d$ pairing model, respectively. Reproduced from Pang \textit{et al.},  Proc. Natl. Acad. Sci. USA $\bm{115}$, 5343 (2018) [Ref. \cite{Pang_2018}].}
	\label{fig5}
\end{figure}

For many years, CeCu$_{2}$Si$_{2}$ had been considered to be a (one-band) $d$-wave superconductor, in particular because of (i) a $T^{3}$ power-law dependence of the spin-lattice relaxation rate and (ii) the absence of a Hebel-Slichter coherence peak at $T_{\rm c}$, owing to Cu-NQR experiments performed down to about 0.1 K \cite{Fujiwara_2008,Ishida_1999}. INS results were compatible with a $d_{x^{2}-y^{2}}$ order parameter \cite{Eremin_2008,Stockert_2011} while (in-plane) magneto-resistivity data favored a $d_{xy}$ state \cite{Viera_2011}. More recently, however, these conclusions were questioned by the detection of a fully developed energy gap below $T \approx 50$~mK as derived from results of high-precision low-$T$ specific-heat measurements \cite{Kittaka_2014}, subsequently supported by the results of heat-conductivity \cite{Yamashita_2017} and new Cu-NQR \cite{Kitagawa_2017} experiments. Figure~\ref{fig5} displays the superfluid density calculated from penetration depth results obtained for an $S$-type CeCu$_{2}$Si$_{2}$ single crystal \cite{Pang_2018}. While no description of the data is possible by a one-band $s$- or $d$- wave model, a fit with two $s$-wave gaps, a large and a small one, is of high quality (Fig.~\ref{fig5}a), see also \cite{Takenaka_2017} $-$ although such a model fails to explain the absence of the Hebel-Slichter peak at $T_{\rm c}$ in the Cu-NQR data. As evident from Fig.~\ref{fig5}b, a similarly good fit of the superfluid-density data (as well as of the specific-heat results by Kittaka \textit{et al.} \cite{Kittaka_2014}) is achieved with a $d + d$ pairing model, alluded to below.

Although the peak in the spin-excitation spectrum below $T_{\rm c}$ should not be confused with the ``spin resonance'' detected for the cuprate high-$T_{\rm c}$ superconductors \cite{Sidis_2004} which shows a negative (``hour-glass'' type) dispersion relation and has been ascribed to a singlet-triplet excitation in the two-particle channel, it nevertheless illustrates a sign-changing superconducting order parameter along the ordering wave vector \bm{$Q_{\rm AF}$} \cite{Bernhoeft_1998,Bernhoeft_2006}. As already mentioned, for CeCu$_{2}$Si$_{2}$ \bm{$Q_{\rm AF}$} $\approx$ \bm{$\tau$}, the theoretically predicted nesting wave vector, connects parallel flat regions within the warped parts of the cylindrical heavy-fermion band, which highlights ``intra-band'' pairing. Alternatively, anisotropic $s$-wave $s_{\pm}$ pairing, which relies on nesting between electron and hole pockets (``inter-band'' pairing) and was successful to describe the gap structure in Fe-based superconductors \cite{Mazin_2008}, has been proposed to explain the finite low-temperature gap in CeCu$_{2}$Si$_{2}$ \cite{Ikeda_2015,Li_2018}. However, as here the ordering wave vector \bm{$Q_{\rm AF}$} is significantly different from any possible inter-band nesting vector connecting the extremely heavy electron band at the $X$-point and the moderately heavy hole bands at the $Z$ point \cite{Wu_2021}, $s_{\pm}$ pairing cannot be applied to CeCu$_{2}$Si$_{2}$. Furthermore, BCS-type isotropic $s$-wave pairing that has also been proposed for CeCu$_{2}$Si$_{2}$ \cite{Takenaka_2017,Yamashita_2017} must be excluded as it cannot explain the peak in the spin-excitation spectrum in the superconducting state. 

The $d + d$ pairing state mentioned before consists of a dominant intra-band $d_{x^{2}-y^{2}}$ and an inter-band $d_{xy}$ component \cite{Nica_2017,Nica_2021}. The intra-band component causes the sign change of the superconducting order parameter inside the warped cylindrical heavy-fermion band. In order to obtain the gap in the zero-temperature limit, these two components have to be added in quadrature which warrants a finite gap value over the whole FS. On the microscopic level, the pairing is equivalent to a matrix-pairing state between the $4f$-electrons in CF-derived $\Gamma_7$ doublets and conduction electrons belonging to $\Gamma_6$ doublets \cite{Nica_2021}. Evidence in favor of this type of pairing was provided by the results of x-ray absorption spectroscopy, which reveal a finite admixture of the $4f$-electron $\Gamma_6$ doublet to the $\Gamma_7$ CF-derived ground-state of CeCu$_{2}$Si$_{2}$ \cite{Amorese_2020}. This $d + d$ pairing theory, which excellently describes the experimental data of the superfluid density displayed in Fig.~\ref{fig5}b, provides the only natural resolution currently available to \textit{all} the observations made on CeCu$_{2}$Si$_{2}$, see \cite{Smidman_2022}.

\begin{figure}[t]
	\begin{center}
		\includegraphics[width=0.5\columnwidth]{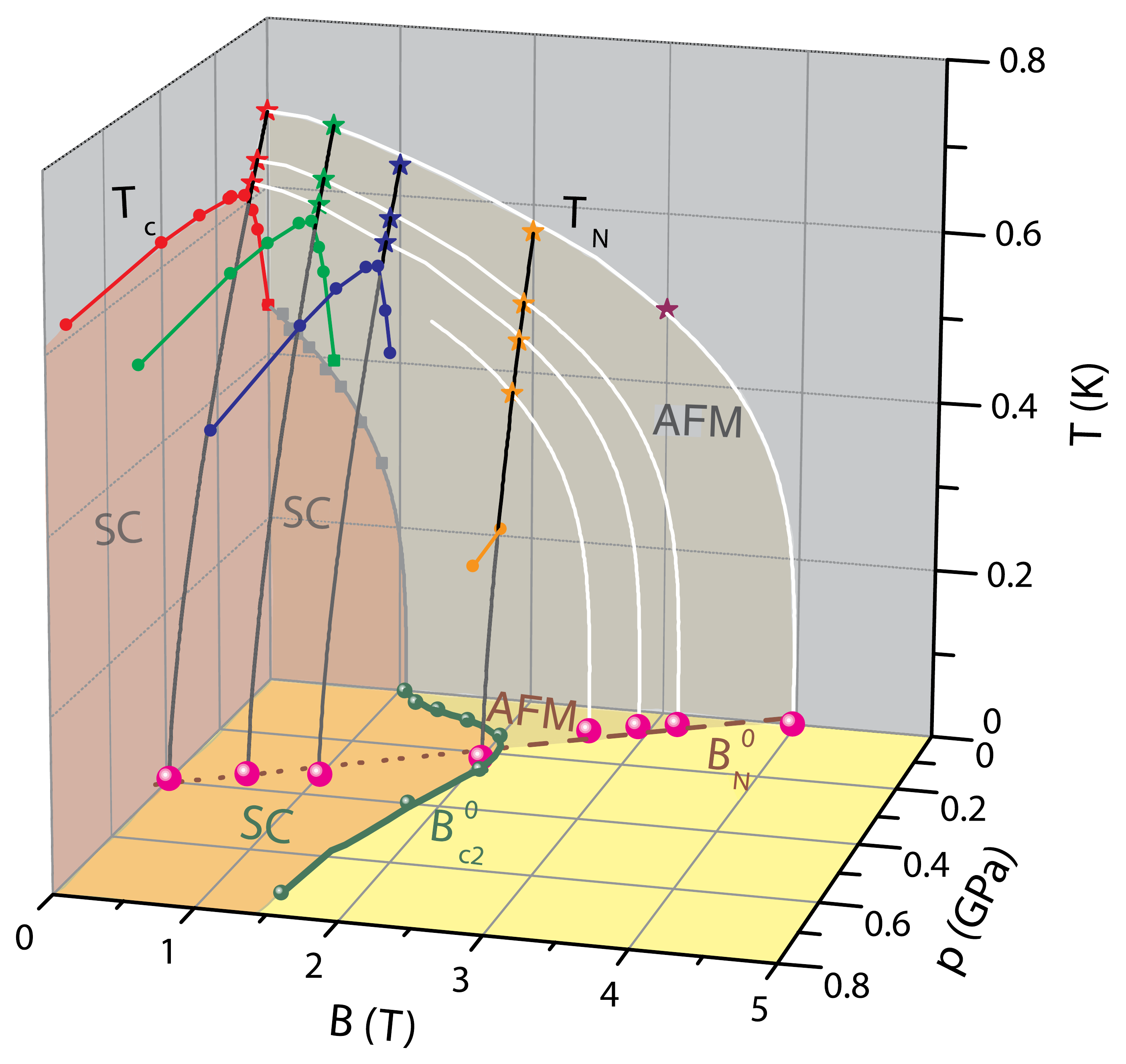}
	\end{center}
	\caption{$T-B-p$ phase diagram of CeCu$_2$Si$_2$, illustrating the zero-temperature line of AF QCPs, intersecting the superconducting dome. Reproduced from Lengyel \textit{et al.}, Phys. Rev. Lett. $\bm{107}$, 057001 (2011) [Ref. \cite{Lengyel_2011}]. Copyright 2011 by the American Physical Society.}
	\label{fig6}
\end{figure}

Turning to the interplay of quantum criticality and SC, one notices that in $A/S$-type CeCu$_{2}$Si$_{2}$ the magnetic phase transition at $T_{\rm N}$ is continuous, whereas the subsequent superconducting transition at lower temperature $T_{\rm c}$ is of $1^{\rm st}$ order, and AF order does not exist as $T \rightarrow 0$ \cite{Feyerherm_1997,Stockert_2006}. By applying a magnetic field in excess of the upper critical field $B_{\rm c2}(T)$, however, the $A$-phase is recovered. The $T-B-p$ phase diagram of an $A/S$-type single crystal is displayed in Fig.~\ref{fig6} \cite{Lengyel_2011}. When extrapolating all the $T_{\rm N}(B)$ resp. $T_{\rm N}(p)$ curves registered at fixed pressure $p$ resp. magnetic-field $B$ to $T = 0$, a line of AF QCPs, $B_{\rm N}^{~0}(p)$, is obtained that can be extrapolated (as $B \rightarrow 0$) to $p_{\rm QC} = 0.39$~GPa. This value is about twice as large as $p_{\rm max}$ where $B_{\rm c2}^{~0}(p)$ intersects the $B_{\rm N}^{~0}(p)$ line and exhibits its maximum. The superconducting dome in the $T = 0$ plane of the phase diagram in Fig.~\ref{fig6} suggests that, upon approaching $p_{\rm QC}$ at low pressure, SC becomes strengthened while it is weakened at elevated $p$, sufficiently close to $p_{\rm QC}$. This indicates that ``Mott-type'' quantum-critical fluctuations between a small and large FS, which are expected to dominate far below $p_{\rm QC}$ (see Fig.~\ref{fig4}a), might be instrumental to form Cooper pairs, as theoretically predicted by Hu \textit{et al.} \cite{Hu_2021}. In contrast, the quantum-critical SDW fluctuations that are dominating in the vicinity of $p_{\rm QC}$ appear to be pair weakening or even pair breaking. Similar conclusions may be drawn from the low-pressure dome, centered around the SDW QCP, in the ($B = 0$) $T-p$ phase diagram of CeCu$_{2}$(Si$_{0.9}$Ge$_{0.1}$)$_{2}$ \cite{Yuan_2003} as well as from related results on other unconventional superconductors showing an SDW-type \cite{Chu_2009} or putative SDW-type \cite{Mathur_1998} QCP.

\section{YbRh$_{2}$Si$_{2}$: QCP at $B \approx 0$ and heavy-fermion superconductivity enabled by nuclear order}

The fact that no SC was detected \cite{Custers_2003} in YbRh$_{2}$Si$_{2}$ above 10~mK is most likely due to the detrimental $4f$-electronic AF spin order which forms at $T_{\rm N} = 70$~mK \cite{Trovarelli_2000}. Likewise, for CeCu$_{2}$Si$_{2}$ SDW order (with $\mu_{\rm s} \leq 0.1 \mu_{\rm B}$) and SC do \textit{not} coexist on a microscopic level \cite{Stockert_2006,Stockert_2011,Feyerherm_1997,Luke_1994}. Interestingly, although AF order is of local-moment nature in YbRh$_{2}$Si$_{2}$ \cite{Gegenwart_2008}, its staggered moment is extremely small, $\mu_{\rm s} \approx 2 \cdot 10^{-3} \mu_{\rm B}$ \cite{Ishida_2003}. On the other hand, \textit{coexistence} of AF order and HF SC has been reported for CeRhIn$_{5}$ \cite{Fobes_2017}, where $\mu_{\rm s}$ is much larger (0.54$\mu_{\rm B}$) \cite{footnote}. Why the size of the staggered moment might influence so strongly the interplay between antiferromagnetism and SC deserves future studies.

\begin{figure}[t]
	\begin{center}
		\includegraphics[width=0.7\columnwidth]{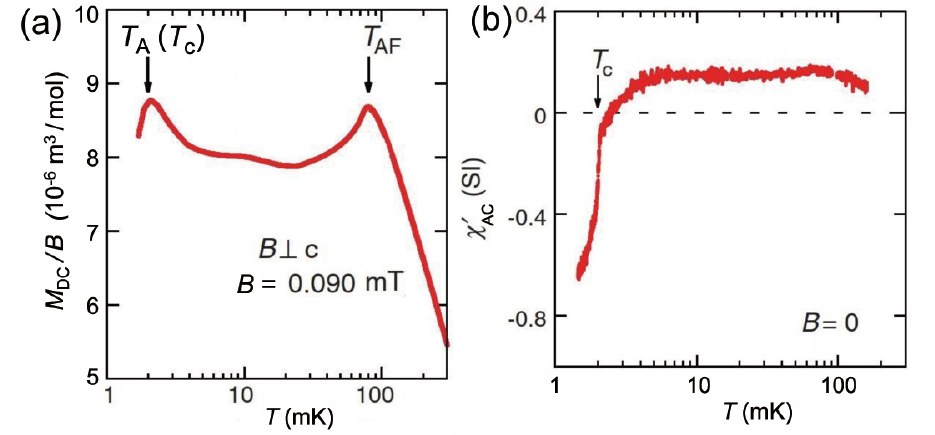}
	\end{center}
	\caption{Temperature dependence of the (a) DC-magnetization, and (b) real part of the AC susceptibility of YbRh$_2$Si$_2$ measured at ultralow temperatures in a nuclear demagnetization cryostat. The positions of the various transitions described in the text are indicated. Note that the peak at $\approx 2$~mK in (a) comprises both the \textit{A}-phase and superconducting transition. Reproduced from Schuberth \textit{et al.}, Science $\bm{351}$, 485 (2016) [Ref. \cite{Schuberth_2016}]. Reprinted with permission from AAAS.}
	
	\label{fig7}
\end{figure}

Using a nuclear demagnetization cryostat with a base temperature of 0.8~mK, Schuberth \textit{et al.} \cite{Schuberth_2016}) discovered HF superconductivity in YbRh$_{2}$Si$_{2}$ at $T_{\rm c} = 2$~mK, slightly below $T_{\rm A} = 2.3$~mK, where a nuclear-dominated hybrid AF order (``\textit{A}-phase'') forms, see the results of field-cooled (fc) magnetization $M_{\rm DC}(T)$ measurements at very low magnetic field presented in Fig.~\ref{fig7}a. The superconducting transition is also displayed by a large diamagnetic shielding signal in the AC susceptibility, $\chi_{\rm AC}(T)$, results shown in Fig.~\ref{fig7}b. Also inferred from this figure is \textit{partial} shielding setting in at about 10~mK which corroborates results of zero-field-cooled (zfc) $M_{\rm DC}(T)$ measurements \cite{Schuberth_2016}. The latter reveal that the shielding is far from being complete just above $T_{\rm c}$. An external magnetic field $B = 23$~mT causes $T_{\rm A}$ to become smaller than 1~mK \cite{Schuberth_2016}. On cooling the sample to below $T \approx 10$~mK, the zfc-magnetization curve is found to split from the fc-curve \cite{Schuberth_2016}. Just below this temperature, the latter shows a marked decrease in its absolute slope (Fig.~\ref{fig7}a).

\begin{figure}[t]
	\begin{center}
		\includegraphics[width=0.65\columnwidth]{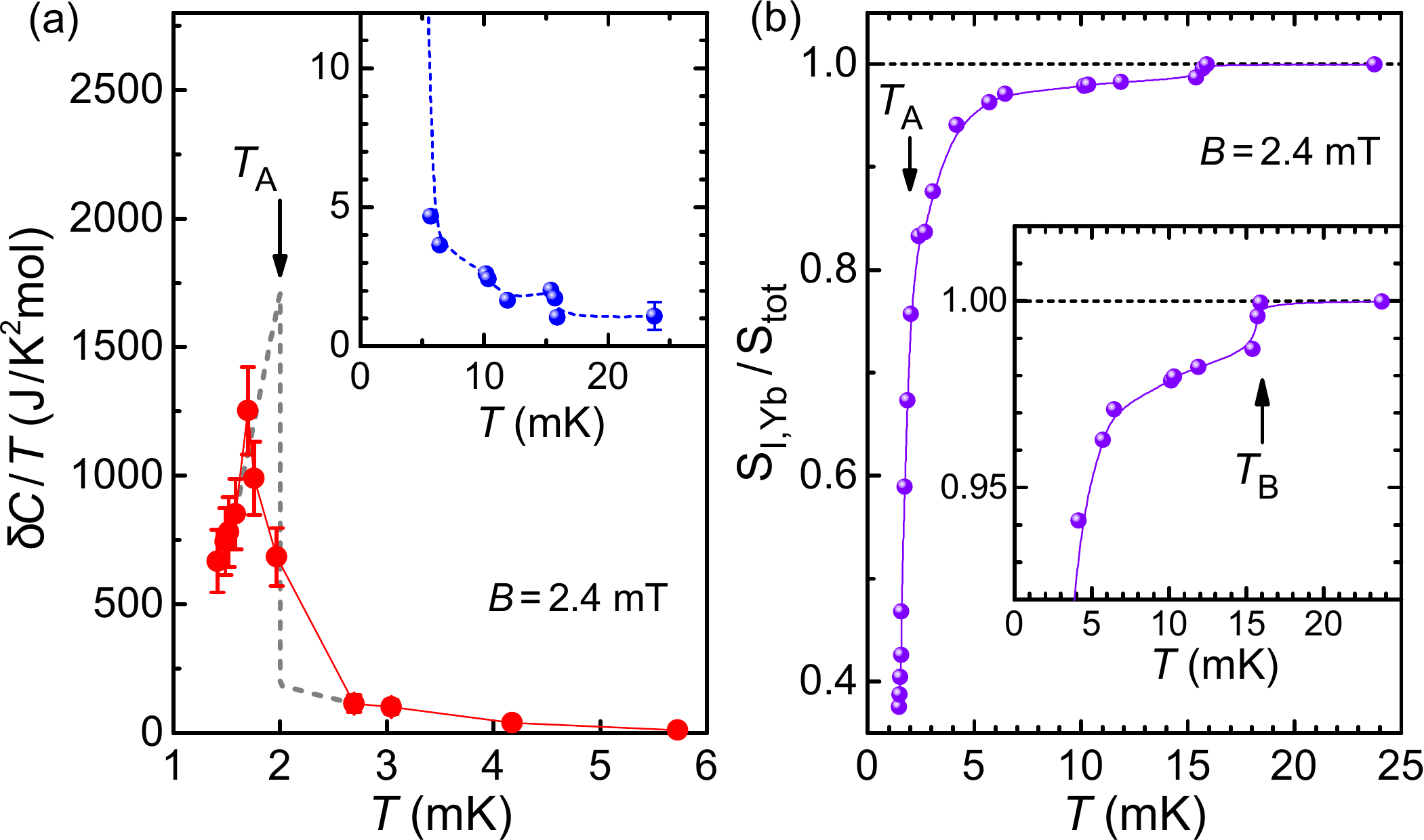}
	\end{center}
	\caption{(a) Heat capacity of YbRh$_2$Si$_2$ measured at ultralow temperatures, demonstrating the transition to the hybrid \textit{A}-phase at $T_{\rm A}$. Inset shows data between 6 mK and 23 mK on a blown-up vertical scale. Adapted from Schuberth \textit{et al.}, Science $\bm{351}$, 485 (2016) [Ref. \cite{Schuberth_2016}]. Reprinted with permission from AAAS. (b) Temperature dependence of the ratio of the entropy originating from the $^{171}$Yb and $^{173}$Yb isotopes to the total entropy in a field of 2.4~mT. Inset highlights the rapid decrease of the nuclear spin entropy at $T_{\rm B} = 16$~mK. Reproduced from Schuberth \textit{et al.}, Front. Electron. Mater. $\bm{2}$: 869495 (2022) [Ref. \cite{Schuberth_2022}]. Available under a Creative Commons Attribution 4.0 International License.}
	
	\label{fig8}
\end{figure}

As displayed in Fig.~\ref{fig8}a, the coefficient $\delta C(T)/T$ of the molar spin specific heat of YbRh$_{2}$Si$_{2}$ (obtained by subtracting a large nuclear-quadrupole contribution from the raw data) reveals a broadened 2$^{\rm nd}$-order phase-transition anomaly which, under conservation of entropy, can be replaced by a gigantic jump (1700~J/K$^{2}$mol), that proves predominant nuclear degrees of freedom \cite{Schuberth_2016}. As seen in the inset, where the data between 6 and 23 mK are presented on a largely expanded vertical scale, $\delta C(T)/T$ undergoes a sharp change at $T_{\rm B} = 16$~mK before reaching the size of the heavy Fermi-liquid value, reported at $B = 0$ below $T = 40$~mK \cite{Gegenwart_2008}. Upon subtracting this $4f$-electronic spin contribution from $\delta C(T)/T$, one can determine the nuclear spin entropy $S_{\rm I,Yb}(T)$ originating from the $^{171}$Yb and $^{173}$Yb isotopes, with nuclear spins $S = 1/2$ and $5/2$, respectively, and an abundance of $\approx 15\%$ in each case \cite{Schuberth_2016}. At $T~\textgreater~16$~mK, $S_{\rm I,Yb}(T)$ apparently assumes its full Zeeman value. While the abrupt decrease of $S_{\rm I,Yb}(T)$ at $T_{\rm B}$ corresponds to  only a very small (1.5$\%$) fraction of Yb nuclear spins being frozen out, this figure increases to about $26\%$ just above $T_{\rm A} = 2.3$~mK (at $B = 0$ \cite{Schuberth_2016}) and to more than 60$\%$ at $\approx 1.5$~mK, see Fig.~\ref{fig8}b.

At $T_{\rm B} = 16$~mK, small droplets of \textit{A}-phase seem to form \cite{Schuberth_2022} which then grow with decreasing temperature and eventually lead to long-range AF order at $T = T_{\rm A}$. In addition, at $T \approx 10$~mK the nucleation of some granular type of SC is inferred from both zfc-$M_{\rm DC}(T)$ \cite{Schuberth_2016} and $\chi_{\rm AC}(T)$ results (Fig.~\ref{fig7}b). This concurs with a significant reduction of the absolute slope of the fc-$M_{\rm DC}(T)$ curve (Fig.~\ref{fig7}a), indicating a (small) Meissner signal \cite{Schuberth_2022}. According to Schuberth \textit{et al.} \cite{Schuberth_2016}, the granular SC is extremely sensitive to the application of tiny magnetic fields. Recent measurements of the electrical resistivity \cite{Nguyen_2021} have provided additional information. While the onset of the resistive transition takes place at $T_{\rm c}^{\rm onset} \approx 9$~mK, zero resistivity is reached at about 6.5~mK \cite{Nguyen_2021}, which highlights percolation of the first superconducting path through the sample \cite{Schuberth_2022}. Interestingly, this percolation goes along with an increase of the \textit{A}-phase short-range order as inferred from a steep rise in the absolute slope of the fc-$M_{\rm DC}(T)$ curve below about 5~mK (Fig.~\ref{fig7}a).

The existence of two subsequent AF phase transitions at $T_{\rm N}$ and $T_{\rm A}$ was treated by R. Yu and Q. Si in the framework of a three-component Landau theory \cite{Schuberth_2016}. This way the primary $4f$-electronic AF phase, which is detrimental to SC, was shown to be suppressed by the dominant nuclear component of the hybrid $A$-phase, establishing a Mott-type AF QCP at (or close to) $B = 0$ and allowing HF SC to form \cite{Schuberth_2022}.

\begin{figure}[t]
	\begin{center}
		\includegraphics[width=0.55\columnwidth]{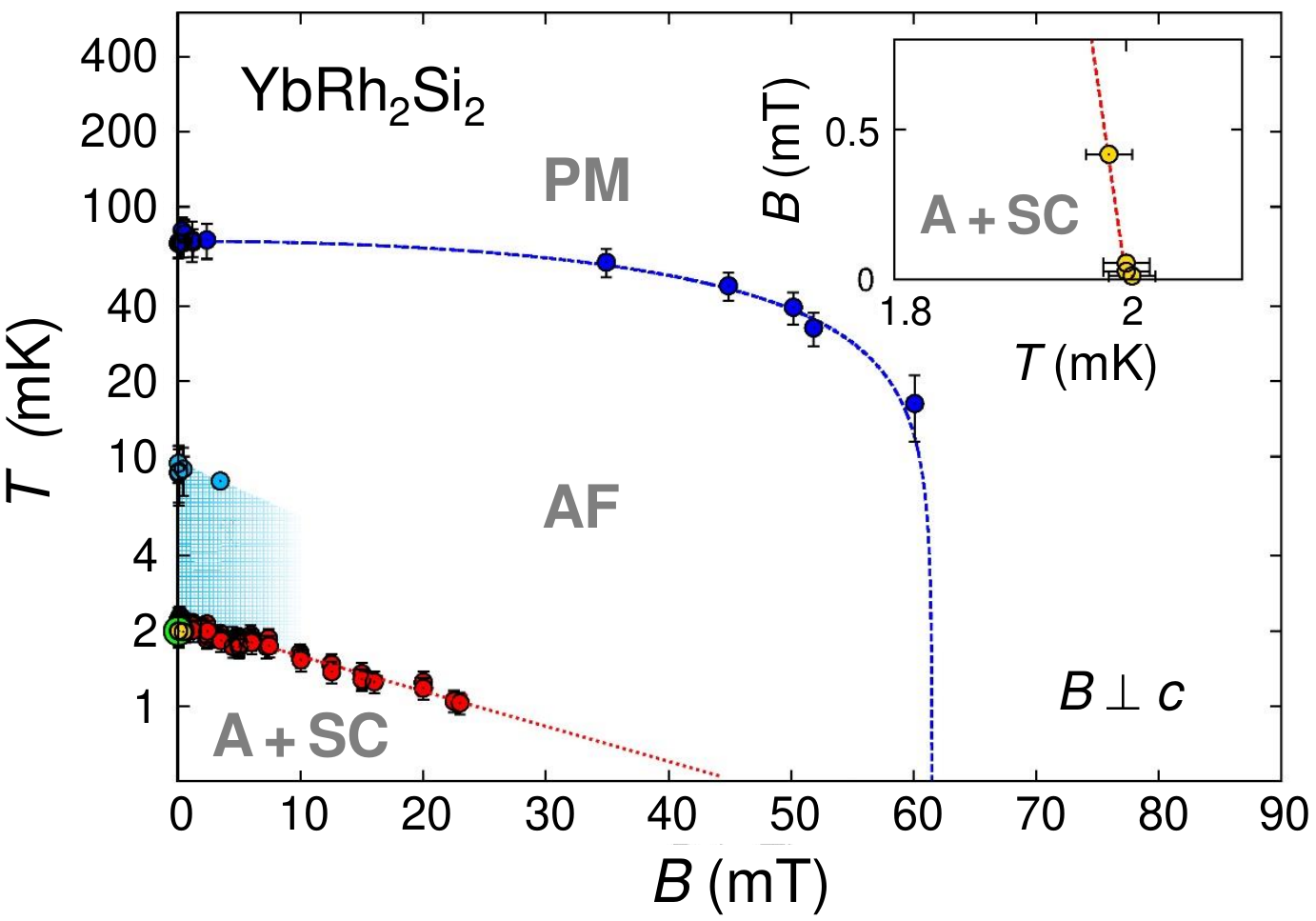}
	\end{center}
	\caption{$T-B$ phase diagram of YbRh${_2}$Si${_2}$ for fields applied perpendicular to the $c$ axis, where the phase boundaries of the paramagnetic (PM), antiferromagnetic (AF), and hybrid superconducting (\textit{A}+SC) phases are marked. The inset displays the field dependence of the jumps in the zfc-$M_{\rm DC}(T)$, yielding an initial slope of $B_{\rm c2}(T)$ at $T_{\rm c}$ of $-25$~T/K, characterising YbRh$_2$Si$_2$ as a HF superconductor.  Adapted from Schuberth \textit{et al.}, Science $\bm{351}$, 485 (2016). [Ref. \cite{Schuberth_2016}]. Reprinted with permission from AAAS.}
	
	\label{fig9}
\end{figure}

The $T-B$ ($B \perp c$) phase diagram of YbRh$_{2}$Si$_{2}$ is shown in Fig.~\ref{fig9}. Inside the primary $4f$-electronic AF order (with $T_{\rm N} = 70$~mK and $B_{\rm N} = 60$~mT), one observes partial (light blue shading) as well as full superconducting shielding (green and yellow data points for $\chi_{AC}$ and zfc-$M_{\rm DC}(T)$, respectively.) The field dependence of the diamagnetic jumps in zfc-$M_{\rm DC}(T)$ presented in the inset yield an initial slope of $B_{\rm c2}(T)$ at $T_{\rm c}$ of $-25$~T/K \cite{Schuberth_2016}, which is typical for a HF superconductor \cite{Assmus1984}. The red dots indicate the (almost) simultaneous onset of \textit{A}-phase ordering and \textit{bulk} SC. The (upper) critical field of the \textit{A}-phase/SC appears to be of the order 40~mT, i.e., about $2/3$ of the quantum critical field $B_{\rm N}$. The ratio of $T_{\rm A} (B = 0) = 2.3$~mK and $B_{\rm A} (T = 0) = 40$~mT determines the effective $g$-factor of the \textit{A}-phase, $g_{\rm eff}\approx0.057$. A comparison with the in-plane $4f$-derived $g$-factor of YbRh$_{2}$Si$_{2}$, $g_{4f} = 3.5$ \cite{Sichelschmidt_2003}, yields a tiny component of $4f$-electronic spins of $\approx 1.6\%$ to the hybrid \textit{A}-phase, the latter being dominated by the contribution of $98.4\%$ of Yb-derived nuclear spins. The mean staggered moment of this $4f$-component amounts to $m_{\rm J} \approx 0.02 \mu_{\rm B}$, which is an order of magnitude larger than the staggered moment of the primary AF phase, $m_{\rm AF} \approx 0.002 \mu_{\rm B}$ \cite{Ishida_2003}. Thus, the $4f$-derived staggered moment substantially increases on cooling to below $T_{\rm A}$ as has also been inferred from noise experiments \cite{Saunders_2018}.

As discussed by Schuberth \textit{et al.} \cite{Schuberth_2022}, the formation of the $4f$-electronic spin component to the hybrid \textit{A}-phase is responsible for about one quarter of the decline observed on cooling to below the low-$T$ peak in fc-$M_{\rm DC}(T)$. Therefore, $\approx 75\%$ of this decline should be attributed to the Meissner effect associated with the bulk superconducting transition \cite{Schuberth_2016,Schuberth_2022}, which takes place at $T_{\rm c} = 2$~mK, slightly below $T_{\rm A} = 2.3$~mK. Interestingly, indications of $granular$ SC at elevated temperatures and fields as concluded from the partial shielding signals alluded to above may also be inferred from the incomplete, step-like resistive transitions reported by Nguyen \textit{et al.} \cite{Nguyen_2021}.

\section{Quantum Critical Paradigm and beyond}

The isostructural HF metals CeCu$_{2}$Si$_{2}$ and YbRh$_{2}$Si$_{2}$ are prototypical Kondo-lattice systems which illustrate the so-called Quantum Critical Paradigm; namely that HF AF QCPs, of either SDW-type like for CeCu$_{2}$Si$_{2}$ or Kondo-destroying-type like for YbRh$_{2}$Si$_{2}$, give rise to unconventional SC. The ultra-low temperature results obtained for YbRh$_{2}$Si$_{2}$ confirm theoretical predictions \cite{Hu_2021} and conclusions drawn from de-Haas-van-Alphen investigations on CeRhIn$_{5}$ \cite{Shishido_2005} that HF SC can indeed form at a ``partial Mott'' QCP. Further work should be devoted to the surprising possibility that the corresponding fermionic quantum critical fluctuations (between small and large FS), which are most likely also responsible for the pronounced ``strange-metal'' behavior of YbRh$_{2}$Si$_{2}$, may be the driving force of HF SC even in the presence of an itinerant SDW QCP with three-dimensional critical fluctuations, like in CeCu$_{2}$Si$_{2}$.

The interplay between HF SC and antiferromagnetism should be scrutinized in more detail, so as to determine why these symmetry-broken phases can coexist if the staggered moment is sufficiently large, whereas SC appears to suppress the AF ordering between small $4f$-moments. Future work is also necessary in order to explore the apparent dichotomy between different measurements of spatial coherence in the Kondo lattice. Owing to transport properties and STS, coherence sets in (Figs.~\ref{fig1}a and \ref{fig2}) at the Kondo temperature $T_{\rm K}$, referring to the CF doublet ground state, but is still not completed even at $T \approx T_{\rm K}/10$ (Figs.~\ref{fig1}b and \ref{fig2}). On the other hand, in ARPES measurements coherence appears to onset at the much higher temperature $T = T_{\rm K}^{\rm high}$, the Kondo temperature corresponding to the fully degenerate Hund's rule multiplet. It is conceivable that issues related to those discussed here for the prototypical Kondo-lattice superconductors CeCu$_{2}$Si$_{2}$ and YbRh$_{2}$Si$_{2}$ will become relevant for other HF metals \cite{Pfleiderer_2009} as well as unconventional superconductors close to a true Mott metal-insulator transistion, such as the cuprates \cite{Lee_2006} and the organic charge-transfer salts \cite{Kanoda_2008}.

\section*{Acknowledgments}
The authors gratefully acknowledge close collaborations and numerous stimulating discussions with Qimiao Si. They are also indebted to Piers Coleman, Gertrud Zwicknagl, Chao Cao, Rong Yu, Emilian Nica, Manuel Brando, Andrea Severing, and Hao Tjeng for helpful conversations.


\begin{thebibliography} {10}
\bibitem{Cornut_1972} B. Cornut, and B. Coqblin, Phys. Rev. B $\bm{5}$, 4541 (1972).
\bibitem{Franz_1978} W. Franz, A. Grie{\ss}el, F. Steglich, and D. K. Wohlleben, Z. Phys. B $\bm{31}$, 7 (1978).
\bibitem{Ocko_2001} M. O\v{c}ko, D. Dobrac, B. Buschinger, C. Geibel, and F. Steglich, Phys. Rev. B $\bm{64}$, 195106 (2001).
\bibitem{Sun_2013} P. Sun, and F. Steglich, Phys. Rev. Lett. $\bm{110}$, 216408 (2013).
\bibitem{Zlatic_2007} V. Zlati\'{c}, R. Monnier, J. K. Freericks, and K. W. Becker, Phys. Rev. B $\bm{76}$, 085122 (2007).
\bibitem{Aliev_1983} F. G. Aliev, N. B. Brandt, V. V. Moshchalkov, and S. M. Chudinov, Solid State Commun. $\bm{47}$, 693 (1983).
\bibitem{Coleman_1985} P. Coleman, P. W. Anderson, and T. V. Ramakrishnan, Phys. Rev. Lett. $\bm{55}$, 414 (1985).
\bibitem{Pikul_2012} A. P. Pikul, U. Stockert, A. Steppke, T. Chichorek, S. Hartmann, N. Caroca-Canales, N. Oeschler, M. Brando, C. Geibel, and F. Steglich, Phys. Rev. Lett. $\bm{108}$, 066405 (2012). 
\bibitem{Kohler_2008} U. K{\"o}hler, N. Oeschler, F. Steglich, S. Maquilon, and Z. Fisk, Phys. Rev. B $\bm{77}$, 104412 (2008).
\bibitem{Ernst_2011} S. Ernst, S. Kirchner, C. Krellner, C. Geibel, G. Zwicknagl, F. Steglich, and S. Wirth, Nature $\bm{474}$, 362 (2011).
\bibitem{Seiro_2018} S. Seiro, L. Jiao, S. Kirchner, S. Hartmann, S. Friedemann, C. Krellner, C. Geibel, Q. Si, F. Steglich, and S. Wirth, Nat. Commun. $\bm{9}$, 3324 (2018).
\bibitem{Zwicknagl_1992} G. Zwicknagl, Adv. Phys. $\bm{41}$, 203 (1992).
\bibitem{Hartmann_2010} S. Hartmann, N. Oeschler, C. Krellner, C. Geibel, S. Paschen, and F. Steglich. Phys. Rev. Lett. $\bm{104}$, 096401 (2010).
\bibitem{Si_2010} Q. Si, and F. Steglich, Science $\bm{329}$, 1161 (2010). 
\bibitem{Si_2001} Q. Si, S. Rabello, K. Ingersent, and J. L. Smith, Nature $\bm{415}$, 804 (2001).
\bibitem{Coleman_2001} P. Coleman, C. P\'{e}pin, Q. Si, and R. Ramazashvili, J. Phys.: Condens. Matter $\bm{13}$, R723 (2001).
\bibitem{Steglich_2014} F. Steglich, H. Pfau, S. Lausberg, S. Hamann, P. Sun, U. Stockert, M. Brando, S. Friedemann, C. Krellner, C. Geibel, S. Wirth, S. Kirchner, E. Abrahams, and Q. Si, J. Phys. Soc. Jpn. $\bm{83}$, 061001 (2014).
\bibitem{Pfau_2012} H. Pfau, S. Hartmann, U. Stockert, P. J. Sun, S. Lausberg, M. Brando, S. Friedemann, C. Krellner, C. Geibel, S. Wirth, S. Kirchner, E. Abrahams, Q. Si, and F. Steglich, Nature $\bm{484}$, 493 (2012).
\bibitem{Pourret_2014} A. Pourret, D. Aoki, M. Boukahil, J.-P. Brison, W. Knafo, G. Knebel, S. Raymond, M. Taupin, Y. \={O}nuki, and J. Flouquet, J. Phys. Soc. Jpn. $\bm{83}$, 061002 (2014).
\bibitem{footnote2} As discussed in Refs. \cite{Steglich_2014,Pfau_2012}, the upturn in $L(T)/{\rm L_{0}}$ below about 100~mK as seen in Fig. \ref{fig3}b is due to a paramagnon contribution to the measured thermal conductivity. This bosonic term has to vanish in the zero-temperature limit, whereby $L(T = 0)/{\rm L_{0}}$ refers to purely electronic transport.
\bibitem{Dong_2013} J. K. Dong, Y. Tokiwa, L. S. Bud’ko, P. C. Canfield, and P. Gegenwart, Phys. Rev. Lett. $\bm{110}$, 176402 (2013).
\bibitem{Schuberth_2022} E. Schuberth, S. Wirth, and F. Steglich, Front. Electron. Mater. $\bm{2}$: 869495 (2022).
\bibitem{Bruin_2013} J. A. N. Bruin, H. Sakai, R. S. Perry, and A. P. Mackenzie, Science $\bm{339}$, 804 (2013).
\bibitem{Taupin_2022} M. Taupin and S. Paschen, Crystals $\bm{12(2)}$, 251 (2022).
\bibitem{Steglich_1975} F. Steglich and J. H. Moeser, in: Low-Temperature Physics LT 14, M. Krusius and M. Vuorio, eds., Vol. 3 (North-Holland; Amsterdam 1975), p. 426. 
\bibitem{Nozieres_1974} P. Nozieres, J. Low Temp. Phys. $\bm{17}$, 31 (1974). 
\bibitem{Chen_2017} Q. Y. Chen, D. F. Xu, X. H. Niu, J. Jiang, R. Peng, H. C. Xu, C. H. P. Wen, Z. F. Ding, K. Huang, L. Shu, Y. J. Zhang, H. Lee, V. N. Strocov, M. Shi, F. Bisti, T. Schmitt, Y. B. Huang, P. Dudin, X. C. Lai, S. Kirchner, H. Q. Yuan, and D. L. Feng, Phys. Rev. B $\bm{96}$, 045107 (2017).
\bibitem{Kummer_2015} K. Kummer, S. Patil, A. Chikina, M. Güttler, M. Höppner, A. Generalov, S. Danzenb{\"a}cher, S. Seiro, A. Hannaske, C. Krellner, Yu. Kucherenko, M. Shi, M. Radovic, E. Rienks, G. Zwicknagl, K. Matho, J. W. Allen, C. Laubschat, C. Geibel, and D. V. Vyalikh, Phys. Rev. X $\bm{5}$, 011028 (2015). 
\bibitem{Stewart_2001and2006} G. R. Stewart, Rev. Mod. Phys. $\bm{73}$, 797 (2001); $\bm{78}$, 743 (2006).
\bibitem{Lohneysen_2007} H. von L{\"o}hneysen, A. Rosch, M. Vojta, and P. W{\"o}lfle, Rev. Mod. Phys. $\bm{79}$, 1015 (2007).
\bibitem{Hertz_1976} J. A. Hertz, Phys. Rev. B $\bm{14}$, 1163 (1976).
\bibitem{Millis_1993} A. J. Millis, Phys. Rev. B $\bm{48}$, 7183 (1993).
\bibitem{Moriya_1995} T. Moriya, and T. Takimoto, J. Phys. Soc. Jpn. $\bm{64}$, 960 (1995). 
\bibitem{Steglich_2016} F. Steglich, and S. Wirth, Rep. Prog. Phys. $\bm{79}$, 084502 (2016). 
\bibitem{Gegenwart_2008} P. Gegenwart, Q. Si, and F. Steglich, Nat. Phys., $\bm{4}$, 186 (2008).
\bibitem{Arndt_2011} J. Arndt, O. Stockert, K. Schmalzl, E. Faulhaber, P. Fouquet, H. S. Jeevan, C. Geibel, W. Schmidt, M. Loewenhaupt, and F. Steglich, Phys. Rev. Lett. $\bm{106}$, 246401 (2011).
\bibitem{Stockert_2011} O. Stockert, J. Arndt, E. Faulhaber, C. Geibel, H. S. Jeevan, S. Kirchner, M. Loewenhaupt, K. Schmalzl, W. Schmidt, Q. Si, and F. Steglich, Nat. Phys. $\bm{7}$, 119 (2011).
\bibitem{Smidman_2022} M. Smidman, O. Stockert, E. M. Nica, Y. Liu, H. Q. Yuan, Q. Si, and F. Steglich, to be published.
\bibitem{Custers_2003} J. Custers, P. Gegenwart, H. Wilhelm, K. Neumaier, Y. Tokiwa, O. Trovarelli, C. Geibel, F. Steglich, C. P\'{e}pin, and P. Coleman, Nature $\bm{424}$, 524 (2003).
\bibitem{Schroder_2000} A. Schr{\"o}der, G. Aeppli, R. Coldea, M. Adams, O. Stockert, H. von L{\"o}hneysen, E. Bucher, R. Ramazashvili, and P. Coleman, Nature $\bm{407}$, 351 (2000).
\bibitem{Shishido_2005} H. Shishido, R. Settai, H. Harima, and Y. \={O}nuki, J. Phys. Soc. Jpn. $\bm{74}$. 1103 (2005). 
\bibitem{Park_2006} T. Park, F. Ronning, H. Q. Yuan, M. B. Salamon, R. Movshovich, J. L Sarrao, and J. D. Thompson, Nature $\bm{440}$, 65 (2006). 
\bibitem{Friedemann_2009} S. Friedemann, T. Westerkamp, M. Brando, N. Oeschler, S. Wirth, P. Gegenwart, C. Krellner, C. Geibel, and F. Steglich, Nat. Phys. $\bm{5}$, 465 (2009). 
\bibitem{Custers_2010} J. Custers, P. Gegenwart, C. Geibel, F. Steglich, P. Coleman, and S. Paschen, Phys. Rev. Lett. $\bm{104}$, 186402 (2010).
\bibitem{Zhao_2019} H. C. Zhao, J. H. Zhang, M. Lv, S. Bachus, Y. Tokiwa, P. Gegenwart, S. Zhang, J. Cheng, Y.-F. Yang, G. F. Chen, Y. Isikawa, Q. Si, F. Steglich, and P. J. Sun, Nat. Phys. $\bm{15}$, 1261 (2019).
\bibitem{Matthias_1958} B. T. Matthias, H. Suhl, and E. Corenzwit, Phys. Rev. Lett. $\bm{1}$, 92 (1958). 
\bibitem{Abrikosov_and_Gorkov_1961} A. A. Abrikosov, and L. P. Gor’kov, Zh. Experim. i. Teor. Fiz. $\bm{39}$, 1781 (1960); JETP $\bm{12}$, 1243 (1961).
\bibitem{Steglich_1979} F. Steglich, J. Aarts, C. D. Bredl, W. Lieke, D. Meschede, W. Franz, and H. Sch{\"a}fer, Phys. Rev. Lett. $\bm{43}$, 1892 (1979). 
\bibitem{Hull_1981} G. W. Hull, J. H. Wernick, T. H. Geballe, J. V. Waszczak, and J. E. Bernadini, Phys. Rev. B $\bm{24}$, 6715 (1981).
\bibitem{Schneider_1983} H. Schneider, Z. Kletowski, F. Oster, and D. Wohlleben, Solid State Commun. $\bm{48}$, 1093 (1983).
\bibitem{Stockert_2004} O. Stockert, E. Faulhaber, G. Zwicknagl, N. St{\"u}{\ss}er, H. S. Jeevan, M. Deppe, R. Borth, R. K{\"u}chler, M. Loewenhaupt, C. Geibel, and F. Steglich, Phys. Rev. Lett. $\bm{92}$, 136401 (2004).
\bibitem{Gegenwart_1998} P. Gegenwart, C. Langhammer, C. Geibel, R. Helfrich, M. Lang, G. Sparn, F. Steglich, R. Horn, L. Donnevert, A. Link, and W. Assmus, Phys. Rev. Lett. $\bm{81}$, 1501 (1998). 
\bibitem{Zwicknagl_1993} G. Zwicknagl, and U. Pulst, Physica B $\bm{186}-\bm{188}$, 895 (1993).
\bibitem{Song_2021} Y. Song, W. Y. Wang, C. D. Cao, Z. Yamani, Y. J. Xu, Y. T. Sheng, W. L{\"o}ser, Y. M. Qiu, Y.-F. Yang, R. J. Birgeneau, P. C. Dai, npj Quantum Materials $\bm{6}$, 60 (2021). 
\bibitem{Sato_2001} N. K. Sato, N. Aso, K. Miyake, R. Shiina, P. Thalmeier, G. Varelogiannis, C. Geibel, F. Steglich, P. Fulde, and T. Komatsubara, Nature $\bm{410}$, 340 (2001).
\bibitem{Song_2016} Y. Song, J. Van Dyke, I. K. Lum, B. D. White, S.-Y. Jang, D.-G. Yazici, L. Shu, A. Schneidewind, P. \v{C}erm\'{a}k, Y. Qiu, M. B. Maple, D. K. Morr, and P. C. Dai, Nat. Commun. $\bm{7}$, 12774 (2016).
\bibitem{Pang_2018} G. M. Pang, M. Smidman, J. L. Zhang, L. Jiao, Z. F. Weng, E. M. Nica, Y. Chen, W. B. Jiang, Y. J. Zhang, W. Xie, H. S. Jeevan, H. Lee, P.~Gegenwart, F. Steglich, Q. Si, and H. Q. Yuan, Proc. Natl. Acad. Sci. USA $\bm{115}$, 5343 (2018).
\bibitem{Fujiwara_2008} K. Fujiwara, Y. Hata, K. Kobayashi, K. Miyoshi, J. Takeuchi, Y. Shimaoka, H. Kotegawa, T. C. Kobayashi, C. Geibel, and F. Steglich, J. Phys. Soc. Jpn. $\bm{77}$, 123711 (2008).
\bibitem{Ishida_1999} K. Ishida, Y. Kawasaki, K. Tabuchi, K. Kashima, Y. Kitaoka, K. Asayama, C. Geibel, and F. Steglich, Phys. Rev. Lett. $\bm{82}$, 5353 (1999).
\bibitem{Eremin_2008} I. Eremin, G. Zwicknagl, P. Thalmeier, and P. Fulde, Phys. Rev. Lett. $\bm{101}$, 187001 (2008).
\bibitem{Viera_2011} H. A. Vieyra, N. Oeschler, S. Seiro, H. S. Jeevan, C. Geibel, D. Parker, and F. Steglich. Phys. Rev. Lett. $\bm{106}$, 207001 (2011).
\bibitem{Kittaka_2014} S. Kittaka, Y. Aoki, Y. Shimura, T. Sakakibara, S. Seiro, C. Geibel, F. Steglich, H. Ikeda, and K. Machida, Phys. Rev. Lett. $\bm{112}$, 067002 (2014).
\bibitem{Yamashita_2017} T. Yamashita, T. Takenaka, Y. Tokiwa, J. A. Wilcox, Y. Mizukami, D. Terazawa, Y. Kasahara, S. Kittaka, T. Sakakibara, M. Konczykowski, S. Seiro, H. S. Jeevan, C. Geibel, C. Putzke, T. Onishi, H. Ikeda, A. Carrington, T. Shibauchi, and Y. Matsuda, Science Adv. $\bm{3}$: e1601667 (2017).
\bibitem{Kitagawa_2017} S. Kitagawa, T. Higuchi, M. Manago, T. Yamanaka, K. Ishida, H. S. Jeevan, and C. Geibel, Phys. Rev. B $\bm{96}$, 134506 (2017).
\bibitem{Takenaka_2017} T. Takenaka, Y. Mizukami, J. A. Wilcox, M. Konczykowski, S. Seiro, C. Geibel, Y. Tokiwa, Y. Kasahara, C. Putzke, Y. Matsuda, A. Carrington, and T. Shibauchi, Phys. Rev. Lett. $\bm{119}$, 077001 (2017).
\bibitem{Sidis_2004} Y. Sidis, S. Pailhes, B. Keimer, P. Bourges, C. Ulrich, and L. P. Regnault, Physica Status Solidi (b) $\bm{241}$, 1204 (2004).
\bibitem{Bernhoeft_1998} N. Bernhoeft, N. Sato, B. Roessli, N. Aso, A. Hiess, G. H. Lander, Y. Endoh, and T. Komatsubara, Phys. Rev. Lett. $\bm{89}$, 4244 (1998). 
\bibitem{Bernhoeft_2006} N. Bernhoeft, A. Hiess, N. Metoki, G. H. Lander, and B. Roessli, J. Phys.: Condens. Matter $\bm{18}$, 5961 (2006).
\bibitem{Mazin_2008} I. I. Mazin, D. J. Singh, M. D. Johannes, and M. H. Du, Phys. Rev. Lett. $\bm{101}$, 057003 (2008).
\bibitem{Ikeda_2015} H. Ikeda, M. T. Suzuki, and R. Arita, Phys. Rev. Lett. $\bm{114}$, 147003 (2015).
\bibitem{Li_2018} Y. Li, M. Liu, Z. Fu, X. Chen, F. Yang, and Y.-F. Yang, Phys. Rev. Lett. $\bm{120}$, 217001 (2018).
\bibitem{Wu_2021} Z. Wu, Y. Fang, H. Su, W. Xie, P. Li, Y. B. Huang, D. W. Shen, B. Thiagarajan, J. Adell, C. Cao, H. Q. Yuan, F. Steglich, and Y. Liu, Phys. Rev. Lett. $\bm{127}$, 067002 (2021).
\bibitem{Nica_2017} E. M. Nica, R. Yu, and Q. Si, npj Quantum Materials $\bm{2}$, 24 (2017). 
\bibitem{Nica_2021} E. M. Nica, and Q. Si, npj Quantum Materials $\bm{6}$, 3 (2021).
\bibitem{Lengyel_2011} E. Lengyel, M. Nicklas, H. S. Jeevan, C. Geibel, and F. Steglich, Phys. Rev. Lett. $\bm{107}$, 057001 (2011).
\bibitem{Amorese_2020} A. Amorese, A. Marino, M. Sundermann, K. Chen, Zh. Hu, T. Willers, F. Choueikani, P. Ohresser, J. Herrero-Martin, S. Agrestini, C.-T. Chen, H.-J. Lin, M. W. Haverkort, S. Seiro, C. Geibel, F. Steglich, L. H. Tjeng, G. Zwicknagl, and A. Severing, Phys. Rev. B $\bm{102}$, 245146 (2020).
\bibitem{Feyerherm_1997} R. Feyerherm, A. Amato, C. Geibel, F. N. Gygax, P. Hellmann, R. H. Heffner, D. E. MacLaughlin, R. Müller-Reisener, G. J. Nieuwenhuys, A. Schenck, and F. Steglich, Phys. Rev. B $\bm{56}$, 699 (1997).
\bibitem{Stockert_2006} O. Stockert, D. Andreica, A. Amato, H. S. Jeevan, C. Geibel, and F. Steglich, Physica B $\bm{374}$-$\bm{375}$, 167 (2006).  
\bibitem{Hu_2021} H. Hu, A. Cai, L. Chen, and Q. Si, arXiv:2109.12794.
\bibitem{Yuan_2003} H. Q. Yuan, F. M. Grosche, M. Deppe, C. Geibel, G. Sparn, and F. Steglich, Science $\bm{302}$, 2104 (2003).
\bibitem{Chu_2009} J.-H. Chu, J. G. Analytis, C. Kucharczyk, and I. R. Fisher, Phys. Rev. B $\bm{79}$, 014506 (2009).
\bibitem{Mathur_1998} N. D. Mathur, F. M. Grosche, S. R. Julian, I. R. Walker, D.M. Freye, R. K. W. Haselwimmer, and G. G. Lonzarich, Nature $\bm{394}$, 39 (1998).
\bibitem{Schuberth_2016} E. Schuberth, M. Tippmann, L. Steinke, S. Lausberg, A. Steppke, M. Brando, C. Krellner, C. Geibel, R. Yu, Q. Si, and F. Steglich, Science $\bm{351}$, 485 (2016).
\bibitem{Trovarelli_2000} O. Trovarelli, C. Geibel, S. Mederle, C. Langhammer, F. M. Grosche, P.  Gegenwart, M. Lang, G. Sparn, and F. Steglich, Phys. Rev. Lett. $\bm{85}$, 626 (2000).
\bibitem{Luke_1994}G. M. Luke, A. Keren, K. Kojima, L. P. Le, B. J. Sternlieb, W. D. Wu, Y. J. Uemura, Y. \={O}nuki, and T. Komatsubara, Phys. Rev. Lett. $\bm{73}$, 1853 (1994).
\bibitem{Ishida_2003} K. Ishida, D. E. MacLaughlin, Ben-Li Young, K. Okamoto, Y. Kawasaki, Y. Kitaoka, G. J. Nieuwenhuys, R. H. Heffner, O. O. Bernal, W. Higemoto, A. Koda, R. Kadono, O. Trovarelli, C. Geibel and F. Steglich, Phys. Rev. B $\bm{68}$, 184401 (2003).
\bibitem{Fobes_2017} D. M. Fobes , E. D. Bauer , J. D. Thompson , A. Sazonov, V. Hutanu, S. Zhang , F. Ronning , and M. Janoschek, J. Phys.: Condens. Matter $\bm{29}$, 17LT01 (2017).
\bibitem{footnote} Also in UPd$_{2}$Al$_{3}$, local-moment ($\mu_{\rm s} = 0.85 \mu_{\rm B}$) AF order coexists with HF SC, but here the U$^{3+}$ ion has a $5f^{3}$ configuration with both localized and itinerant $5f$ electrons \cite{Sato_2001,Zwicknagl_2003}.
\bibitem{Nguyen_2021} D. H. Nguyen, A. Sidorenko, M. Taupin, G. Knebel, G. Lapertot, E. Schuberth, and S. Paschen, Nat. Commun. $\bm{12}$, 4341 (2021).
\bibitem{Assmus1984} W. Assmus, M. Herrmann, U. Rauchschwalbe, S. Riegel, W. Lieke, H. Spille, S. Horn, G. Weber, F. Steglich, and G. Cordier, Phys. Rev. Lett. $\bm{52}$, 469 (1984).
\bibitem{Sichelschmidt_2003} J. Sichelschmidt, V. A. Ivanshin, J. Ferstl, C. Geibel, and F. Steglich, Phys. Rev. Lett. $\bm{91}$, 156401 (2003).
\bibitem{Saunders_2018} J. Saunders, \textit{``Superconductivity in YbRh$_{2}$Si$_{2}${\rm :} Electrical Transport and Noise Experiments,''} Invited Talk at 12$^{\rm th}$ Intern. Conf. on Materials and Mechanisms of Superconductivity (M$^{2}$S 2018), Beijing, China, August 19–24 (2018). 
\bibitem{Pfleiderer_2009} C. Pfleiderer, Rev. Mod. Phys. $\bm{81}$, 1551 (2009).
\bibitem{Lee_2006} P. A. Lee, N. Nagaosa, and X.-G. Wen, Rev. Mod. Phys. $\bm{78}$, 17 (2006). 
\bibitem{Kanoda_2008} K. Kanoda, in: The Physics of Organic Superconductors and Conductors, A. Lebed (ed.), (Berlin, Heidelberg, Springer-Verlag), p. 623 (2008).
\bibitem{Zwicknagl_2003} G. Zwicknagl, A. Yaresko, and P. Fulde, Phys. Rev. B $\bm{68}$, 052508 (2003).




\end{thebibliography}
\end{document}